\documentclass{article}
\usepackage{spconf,amsmath,graphicx}

\usepackage[dvipsnames]{xcolor}
\usepackage{booktabs}
\usepackage{multirow}
\usepackage{amsmath}
\usepackage{amsfonts}
\usepackage{hyperref}
\usepackage[T1]{fontenc}

\title{fine-grained emotional control of text-to-speech: \\ learning to rank inter- and intra-class emotion intensities }
\name{Shijun Wang$^1$, Jón Guðnason$^2$, Damian Borth$^1$}
\address{
  $^1$University of St.Gallen, Switzerland\\
  $^2$Reykjavik University, Iceland}
\begin{document}
%
\maketitle
\begin{abstract}

State-of-the-art Text-To-Speech (TTS) models are capable of producing high-quality speech. 
The generated speech, however, is usually neutral in emotional expression, whereas very often one would want fine-grained emotional control of words or phonemes. Although still challenging, the first TTS models have been recently proposed that are able to control voice by manually assigning emotion intensity.
Unfortunately, due to the neglect of intra-class distance, the intensity differences are often unrecognizable.
In this paper, we propose a fine-grained controllable emotional TTS, that considers both inter- and intra-class distances and be able to synthesize speech with recognizable intensity difference.
Our subjective and objective experiments demonstrate that our model exceeds two state-of-the-art controllable TTS models for controllability, emotion expressiveness and naturalness.
\end{abstract}
\begin{keywords}
emotional TTS, emotion intensity control, speech emotion analysis
\end{keywords}
\section{Introduction}
\label{sec:intro}
\vspace{-0.3cm}
Recent end-to-end Text-To-Speech (TTS) models \cite{Sotelo2017Char2WavES, Shen2018NaturalTS, Chung2021ReinforceAlignerRA, Ren2021FastSpeech2F, Tan2021ASO} have the capacity to synthesize high-quality speech with neutral emotion.
These models are, however, limited when it comes to expressing paralinguistic information such as emotion.
It is critical to address this issue because expressing emotion in speech is crucial in many applications such as audiobook generation or digital assistants.
Moreover, an additional challenge of current TTS models is the lack of fine-grained controllability of emotion on words or phonemes.
Such a drawback results in inflexible control of speech, and failure to meet the context or users' intentions.

One straightforward strategy to express different emotions is by conditioning global emotion labels \cite{Lee2017EmotionalEN, Neekhara2021ExpressiveNV}. 
However, synthesized speech from these models has monotonous emotional expression due to the condition of one global emotion representation. 
To achieve diverse emotion expression, models like GST \cite{Wang2018StyleTU} apply a token (a single vector) to represent the emotional style of a reference speech, then use this token to influence the synthesis.
RFTacotron \cite{Lee2019RobustAF} is an extended work of GST. It uses a sequence of vectors instead of a single token to represent the emotion, which allows the improvement of the robustness and prosody control. 
Nevertheless, the nuance of references might be difficult to be captured by these models (e.g. one sad and one depressed reference might produce the same synthesized speech), due to a mismatch between the content or speaker of the reference and synthesized speech, which implies the inflexible controllability of these models. 

A better approach to achieve fine-grained controllable emotional TTS is by manually assigning intensity labels (such as strong or weak happiness) on words or phonemes, which provides a flexible and efficient way to control the emotion expression, even for subtle variations.
In \cite{Zhu2019ControllingES, lei2021fine, lei2022msemotts, schnell2022controllability}, Rank algorithms are used to extract emotion intensity information, by following the assumptions: i) speech samples from the same emotion class have similar ranks, and ii) intensity of neutral emotion is the weakest and all other emotions are ranked higher than neutral.
Despite the production of recognizable speech samples with different emotion intensity levels, intra-class distance is neglected in these models. 
Specifically, during the training, samples belonging to the same emotion class (for instance, the strongest and weakest happiness) are arbitrarily considered the same. 
In practice, confusion could happen when we compare a median-level intensity speech with a strong- or weak-level one.

In this paper, we propose a TTS model, which outperforms the state-of-the-art fine-grained controllable emotional TTS models.  
The model is based on a novel Rank model, which is simple yet efficient for extracting emotion intensity information, by taking into account both inter- and intra-distance.
Instead of performing rank on a non-neutral and a neutral sample, we use two samples augmented by Mixup \cite{Zhang2018mixupBE}.
Each augmented sample is a mixture from the same non-neutral and neutral speech.
By applying different weights to non-neutral and neutral speech, one mixture contains more non-neutral components than the other one. In other words, one mixture's non-neutral intensity is stronger than that of the other. By learning to rank these two mixtures, our Rank model not only needs to determine the emotion class (inter-class distance), but also has to capture the amount of non-neutral emotion present in a mixed speech, i.e. intensity of non-neutral emotion  (intra-class distance).

We summarize our contributions as: 
1) we propose a fine-grained controllable emotional TTS model based on a novel Rank model.
2) The proposed Rank model is simple and efficient to extract intensity information.
3) Our experimental results demonstrate that our TTS model outperforms two state-of-the-art fine-grained controllable emotional TTS models.
Demo page can be found at \url{https://wshijun1991.github.io/ICASSP2023\_DEMO/}.

\vspace{-0.3cm}
\section{Approach}
\label{sec:approach}
\vspace{-0.3cm}
We train two models.
One is a Rank model that aims to extract emotion intensity representations.
The other is a backbone TTS model used to generate speech.

\vspace{-0.3cm}
\subsection{Rank Model}
\label{sec:Rank}
\begin{figure}[t]
  \centering
  \includegraphics[width=0.65\linewidth]{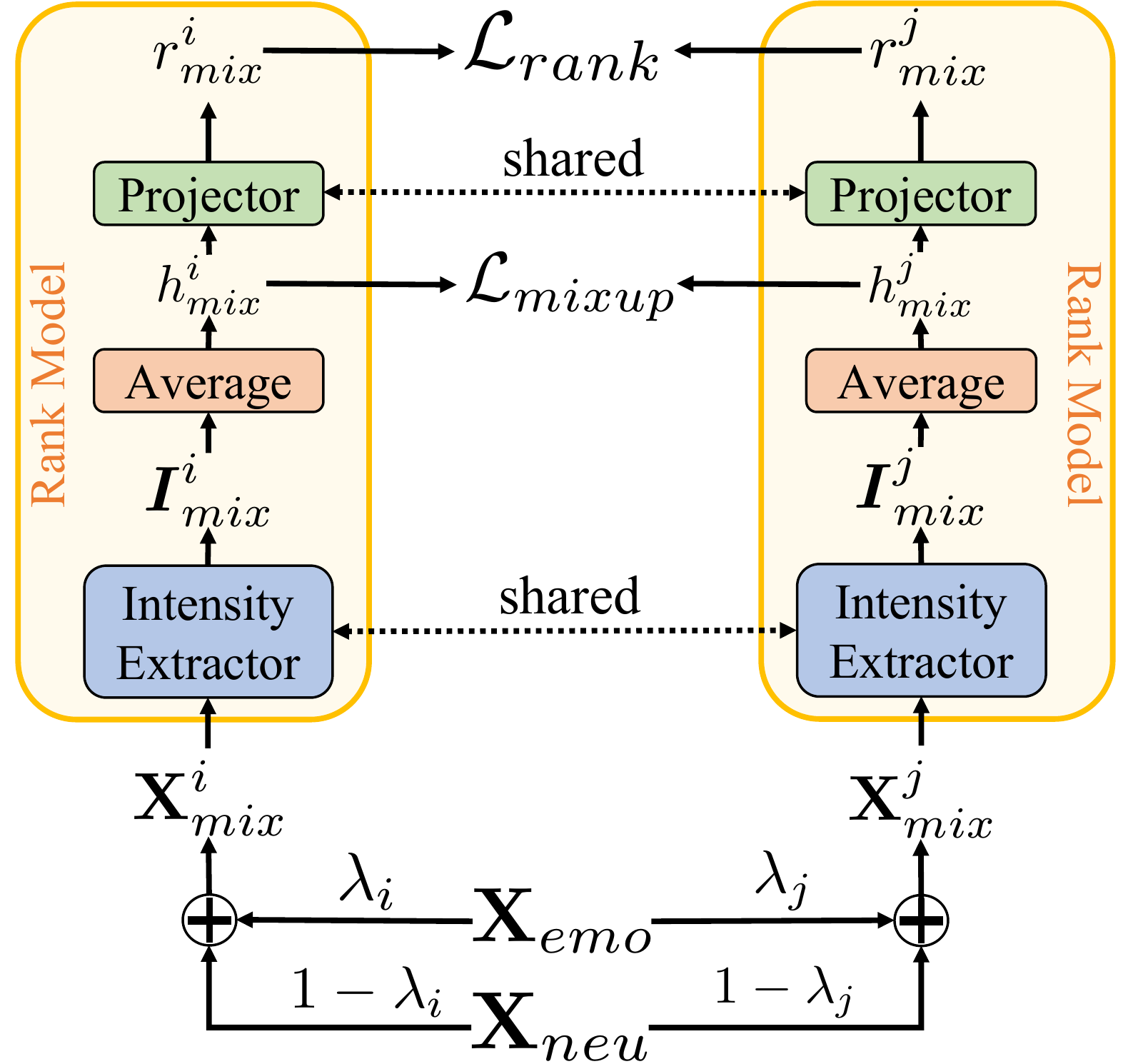}
  \caption{The training of our Rank model. $\mathbf{X}_{emo}$ is a speech sample of non-neutral emotion classes, and $\mathbf{X}_{neu}$ is neutral. 
  The Intensity Extractor produces intensity representations $\boldsymbol{I}_{mix}^i$ and $\boldsymbol{I}_{mix}^j$ given mixtures $\mathbf{X}_{mix}^i$ and $\mathbf{X}_{mix}^j$.
  $h_{mix}^i$ and $h_{mix}^j$ are two averaged vectors.
  $\mathcal{L}_{mixup}$ is a weighted cross entropy loss, while $\mathcal{L}_{rank}$ is to rank $r_{mix}^i$ and $r_{mix}^j$, two scores, regarding the intensity of non-neutral emotion.}
  \label{rank_fig}
  \vspace{-0.3cm}
\end{figure}

\vspace{-0.2cm}
Our Rank model is shown in Fig. \ref{rank_fig}. It maps the speech into intensity representations, then outputs a rank score regarding the emotion intensity.
Input $\mathbf{X}$ is a concatenation of Mel-Spectrogram, pitch contour, and energy. $\mathbf{X}_{neu}$ indicates an input from neutral class, while $\mathbf{X}_{emo}$ represents an input from other non-neutral emotion classes.
We then perform Mixup augmentation on the pair ($\mathbf{X}_{neu}$, $\mathbf{X}_{emo}$):
\begin{equation}
\begin{aligned}
  \mathbf{X}_{mix}^i &= \lambda_i \mathbf{X}_{emo} + (1-\lambda_i) \mathbf{X}_{neu}, \\
  \mathbf{X}_{mix}^j &= \lambda_j \mathbf{X}_{emo} + (1-\lambda_j) \mathbf{X}_{neu}, 
\end{aligned}
\end{equation}
where $\lambda_i$ and $\lambda_j$ are from Beta distribution $Beta(1,1)$.

The Intensity Extractor is then used to extract intensity representations. It first applies the same Feed-Forward Transformer (FFT) in \cite{Ren2021FastSpeech2F} to process the input.
We further add an emotion embedding to the output of FFT to produce intensity representations $\boldsymbol{I}_{mix}^i$ and $\boldsymbol{I}_{mix}^j$. This embedding is from a look-up table and depends on the emotion class of $\mathbf{X}_{emo}$.
The addition of emotion embedding is to provide information on emotion class, because intensity might vary differently in various emotion classes.

From the intensity representations $\boldsymbol{I}_{mix}^i$ and $\boldsymbol{I}_{mix}^j$, we then average these two sequences into two vectors $h_{mix}^i$ and $h_{mix}^j$.
The original Mixup loss is further applied on them:
\begin{equation}
\begin{gathered}
\mathcal{L}_{mixup} = \mathcal{L}_{i} + \mathcal{L}_{j}, \text{where} \\
\mathcal{L}_{i} = \lambda_i \text{CE}(h_{mix}^i, y_{emo}) + (1-\lambda_i) \text{CE}(h_{mix}^i, y_{neu}), \\
\mathcal{L}_{j} = \lambda_j \text{CE}(h_{mix}^j, y_{emo}) + (1-\lambda_j) \text{CE}(h_{mix}^j, y_{neu}), 
\end{gathered}
\end{equation}
and CE($\cdot, \cdot$) represents Cross Entropy loss, $y_{emo}$ indicates labels for non-neutral emotion, while $y_{neu}$ indicates neutral.

Despite the fact that Mixup has been demonstrated as an effective regularization method, there is little evidence showing it is sensitive to the intra-class distance. 
Thus, apart from $\mathcal{L}_{mixup}$ (inter-class), we need to introduce another loss to capture intra-class information.
Inspired by \cite{wang2022zero}, we first use a Projector  (linear layers) to map the pair ($h_{mix}^i$, $h_{mix}^j$) to a scalar pair ($r_{mix}^i$, $r_{mix}^j$), where $r_{mix} \in \mathbb{R}^1$.
$r_{mix}$ is a score indicating the amount of non-neutral emotion present in speech, i.e. intensity.
To force the model to correctly assign scores, we first feed the score difference into a Sigmoid function:
\begin{equation}
p^{ij} = \frac{1}{1+e^{-(r_{mix}^i-r_{mix}^j)}},
\label{eq3}
\end{equation}
then we apply the rank loss on it:
\begin{equation}
\mathcal{L}_{rank} = -\lambda_{diff} \text{log}(p^{ij}) - (1-\lambda_{diff}) \text{log}(1-p^{ij}),
\label{eq4}
\end{equation}
where $\lambda_{diff}$ is a normalized result of $\lambda_i-\lambda_j$, which means if $\lambda_i > \lambda_j$, then $\lambda_{diff} \in (0.5, 1)$; if $\lambda_i < \lambda_j$, then $\lambda_{diff} \in (0, 0.5)$; if $\lambda_i = \lambda_j$, then $\lambda_{diff}=0.5$. 

As an example, if $\lambda_i > \lambda_j$ (non-neutral emotion presents more in $\mathbf{X}_{mix}^i$ compared to $\mathbf{X}_{mix}^j$), 
then $\lambda_{diff} > 0.5$,
in this case, in order to decrease the rank loss in Eq. \ref{eq4}, the model needs to assign a bigger $r_{mix}^i$ for $\mathbf{X}_{mix}^i$, to enable the Sigmoid output in Eq. \ref{eq3} to be bigger than 0.5.

The intuition is forcing the model to correctly rank two samples that both contain non-neutral emotion. 
To achieve this, the intensity representation $\boldsymbol{I}_{mix}$ must convey information that can indicate the intensity of non-neutral emotion.

Lastly, we train our Rank model with the total loss:
\begin{equation}
\mathcal{L}_{total} = \alpha \mathcal{L}_{mixup} + \beta \mathcal{L}_{rank},
\label{eq5}
\end{equation}
where $\alpha$ and $\beta$ are the loss weights.

\begin{figure}[t]
  \centering
  \includegraphics[width=0.7\linewidth]{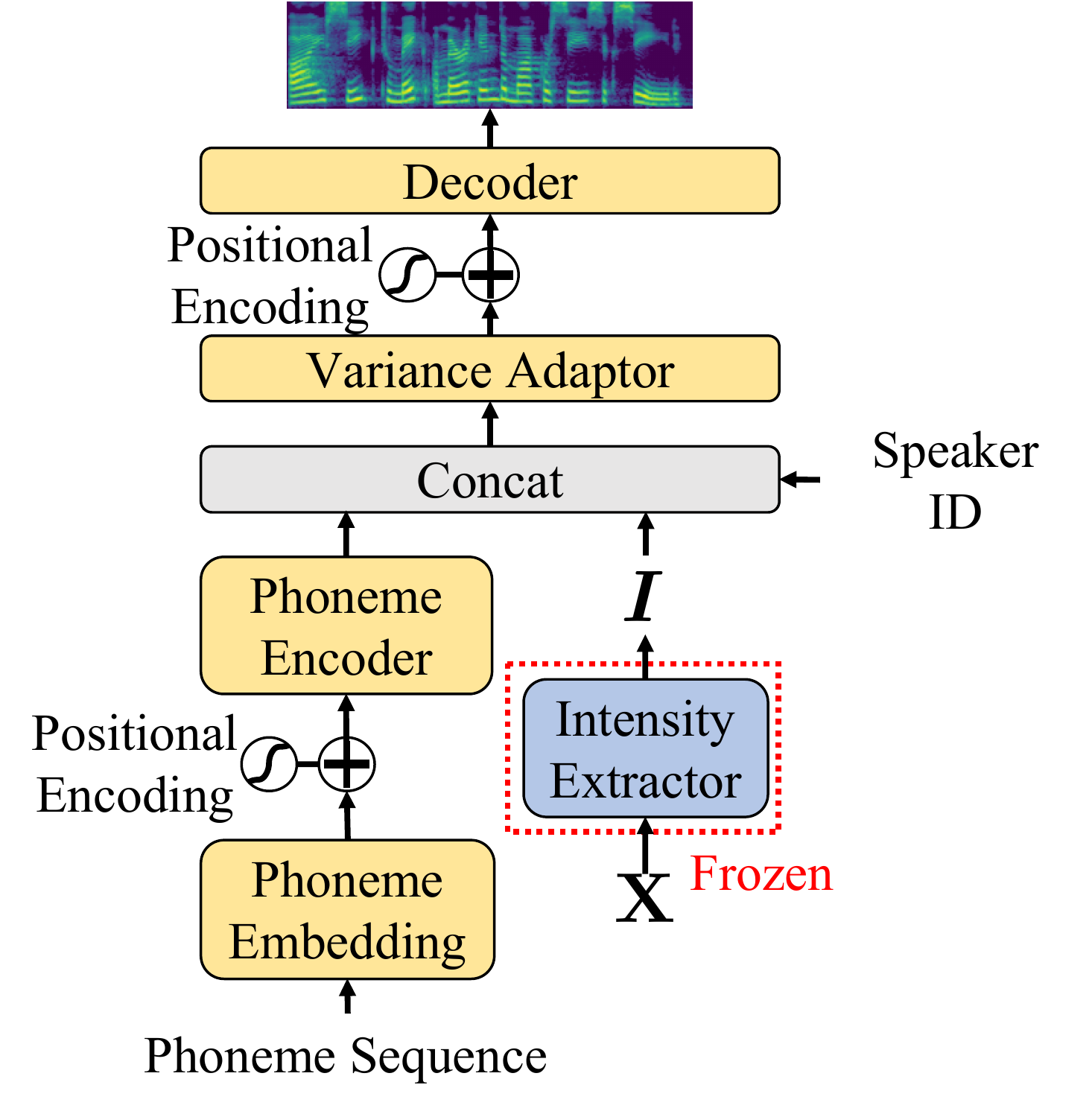}
  \caption{The training of FastSpeech2, a trained Intensity Extractor is combined to provide intensity representations.}
  \label{tts_fig}
  \vspace{-0.3cm}
\end{figure}

\vspace{-0.3cm}
\subsection{TTS Model}
\vspace{-0.2cm}
We use FastSpeech2 \cite{Ren2021FastSpeech2F} to convert the phonemes to speech, given intensity information.
We maintain the original model configuration, except our Intensity Extractor is combined to provide intensity information.

The training of FastSpeech2 is shown in Fig. \ref{tts_fig}, and we only give a short description of each module here but refer the readers to the original paper \cite{Ren2021FastSpeech2F} for more in-depth description.
The Phoneme Encoder is to process phoneme and position information. 
The speaker ID is mapped to speaker embedding to represent speaker characteristics.
The Variance Adaptor aims to predict pitch, energy and duration (frame length of each phoneme) based on the input.
The decoder generates the final Mel-Spectrogram.

To incorporate intensity information, a pre-trained Intensity Extractor is frozen and integrated.
And the Variance Adaptor uses phoneme, speaker information, and intensity representation $\boldsymbol{I}$ to predict pitch, energy, and duration. 
We set intensity representations for neutral emotion to zero, since we assume there is no intensity variation for neutral speech.
One thing to point is that since the length of $\boldsymbol{I}$ is not equal to the phoneme length, we use Montreal Forced Aligner \cite{McAuliffe2017MontrealFA} to acquire intensity segments corresponding to each phoneme. Then the intensity segments are averaged to make the lengths of intensity representation and phoneme the same.

\vspace{-0.3cm}
\subsection{Training and Inference}
\vspace{-0.2cm}
\noindent 
\textbf{Training}: We first train our Rank model. Then the Intensity Extractor from the trained Rank model is frozen and combined during the training of FastSpeech2.

\noindent 
\textbf{Inference}: During inference, we expect to use phonemes and manual intensity labels to control the emotion intensity of synthesized speech. 
However, our Intensity Extractor can only output intensity representation from speech.
In order to achieve controlling intensity with manual labels, we use the following strategy: 
with a trained Intensity Extractor, we first collect all intensity representations and their intensity scores. 
Then, we bucket all scores into several bins, where each bin denotes one intensity level (e.g. in our work, we use Min, Median and Max intensity levels, which means we apply three bins).
After that, intensity representations corresponding to each intensity level are averaged into a single vector. 
Finally, we can map manual intensity labels to intensity representations during inference. 
This strategy is applied to each emotion class, therefore, we can find individual intensity representations by feeding emotion and manual intensity labels.

\noindent
\textbf{Implementation Details}: We train the Rank model for 20k iterations with 1e-6 learning rate. For FastSpeech2, we use 250k iterations with 1e-4 learning rate. Adam optimizer is applied for both cases. In Eq. \ref{eq5}, $\alpha$ and $\beta$ are respectively set as 0.1 and 1.0.
\vspace{-0.3cm}
\section{Experiments}
\label{sec:Experiments}
\vspace{-0.2cm}

\subsection{Experimental Setup}
\vspace{-0.2cm}
\noindent 
\textbf{Dataset}: EmoV-DB \cite{Adigwe2018TheEV} is used as our dataset, it contains four speakers and five emotions (Amused, Angry, Disgusted, Neutral and Sleepy). The overall speech samples are around 7 hours and the sampling rate is 16KHz. 

\noindent 
\textbf{Data Preprocessing}: Since FastSpeech2 is used as our backbone TTS, we need to feed Mel-Spectrogram, pitch and energy as inputs. 
We use 50-millisecond window and a
50 percent overlap ratio to extract energy and Mel-Spectrogram with 80 Mel Coefficients. PyWorld\footnote{https://github.com/JeremyCCHsu/Python-Wrapper-for-World-Vocoder} is applied to extract pitch.

\noindent 
\textbf{Baselines}: We use FEC \cite{lei2021fine} and RFTacotron \cite{Lee2019RobustAF} as our baselines. 
Similar to our model, FEC has a Rank model and allows you to assign emotion intensity to each phoneme as well. To ensure a fair comparison, we replace the original Tacotron2 \cite{Shen2018NaturalTS} in FEC with our FastSpeech2.
RFTacotron transfers emotional style from a reference into the synthesized speech, which can be used to control the emotion intensity by applying reference samples with different intensities. We keep using the original Tacotron2 because its attention mechanism is the key part for emotion transfer.

\noindent 
\textbf{Evaluation Setup}: We use PWGAN \cite{Yamamoto2020ParallelWA} to convert generated Mel-Spectrograms to waveforms.
To perform objective and subjective evaluations, for each emotion and speaker, we randomly select 5 unseen speech samples, by using their corresponding utterances, 90 utterances (speaker Josh has no data for Angry and Disgusted) are prepared for evaluation. 
20 subjects participate in the subjective evaluations.

\begin{table}[t]
  \small
  \caption{Rate Accuracy of Emotion Intensities. Min, Median and Max are three intensity levels. Subjects are asked to select the sample with the stronger intensity from a pair.}
  \label{intensity_rate}
  \centering
    \begin{tabular}{ c | c | c | c | c }
    \toprule
    \multirow{3}*{Emotion}    & \multirow{3}*{Models}     & \multicolumn{3}{c}{Intensity Pairs}            \\ 
    \cline{3-5}
                              &                           & Min           & Median       &Min              \\
                              &                           & -Median   & -Max     &-Max         \\

    \midrule
    \multirow{3}*{Amused}     & RFTacotron                & 0.38            & 0.52           & 0.59              \\ 
                              & FEC                       & 0.63            & 0.58           & 0.63              \\ 
                              & Ours                      & \textbf{0.66}            & \textbf{0.60}           & \textbf{0.74}              \\ 
    \midrule

    \multirow{3}*{Angry}      & RFTacotron                & 0.53            & 0.59           & 0.60              \\ 
                              & FEC                       & 0.59            & 0.58           & 0.73              \\ 
                              & Ours                      & \textbf{0.65}            & \textbf{0.67}           & \textbf{0.75}              \\ 
    \midrule

    \multirow{3}*{Sleepy}     & RFTacotron                & 0.39            & 0.45           & 0.52              \\ 
                              & FEC                       & 0.56            & 0.54           & 0.64              \\ 
                              & Ours                      & \textbf{0.65}            & \textbf{0.73}           & \textbf{0.83}              \\ 
    \midrule

    \multirow{3}*{Disgusted}  & RFTacotron                & 0.48            & 0.51           & 0.53              \\ 
                              & FEC                       & 0.57            & 0.63           & 0.72              \\ 
                              & Ours                      & \textbf{0.72}            & \textbf{0.67}           & \textbf{0.75}              \\ 
    \midrule
    \midrule
    \multirow{3}*{Average}    & RFTacotron                & 0.45            & 0.52           & 0.56              \\ 
                              & FEC                       & 0.61            & 0.58           & 0.68              \\ 
                              & Ours                      & \textbf{0.67}            & \textbf{0.67}           & \textbf{0.77}              \\

    \bottomrule
    \end{tabular}
    \vspace{-0.2cm}
\end{table}

\begin{table}[t]
  \small
  \caption{MCD and Naturalness MOS}
  \label{mcd}
  \centering
    \begin{tabular}{ c | c | c}
    \toprule
    Model                     & MCD (dB)   &MOS             \\ 
    \midrule
    Ground Truth    &/ & 3.9 ± 0.05 \\

    \midrule
    RFTacotron                & 5.21    &3.49 ± 0.04
             \\ 
    FEC                       & 4.79   & 3.7 ± 0.04
         \\ 
    Ours                      & \textbf{4.66}   & \textbf{3.76 ± 0.03}
         \\

    \bottomrule
    \end{tabular}
    \vspace{-0.2cm}
\end{table}

\vspace{-0.3cm}
\subsection{Emotion Intensity Controllability}
\vspace{-0.2cm}
In this section, we perform subjective evaluation to detect how recognizable of synthesized speech samples with different intensities (Min, Median or Max).
Easily recognizable synthesized speech samples imply that we can efficiently control the emotion intensity by manually assigning intensity labels.
Like in \cite{Sivaprasad2021EmotionalPC}, for each utterance, we first synthesize three speech samples with three intensity levels, respectively. Then, three pairs (Min-Max, Min-Median and Median-Max) can be acquired and we ask subjects to select which one from a pair contains a stronger intensity.
If the one that selected by subjects is the one synthesized with a stronger intensity, then it shows the emotion intensity can be appropriately controlled. 

In FEC, the intensity rank scores are normalized in $(0,1)$, thus, we refer to scores in $(0,0.33]$ as Min, $(0.33,0.66]$ as Median, and $(0.66,1.0)$ as Max. 
Since RFTacotron is unable to assign intensity scores, we use our Rank model (Sec. \ref{sec:Rank}) to find the strongest, median and the weakest samples from the dataset as references (based on the rank score $r$).
We believe it is a fair comparison, because if our Rank model fails, then both RFTacotron and our model should fail.

As we can see from the results (Tab. \ref{intensity_rate}), RFTacotron might not be efficient for performing intensity control, in some cases, the intensity difference is not easily perceivable. 
FEC improves a lot regarding the control of intensity, however, confusion happens when a median-level sample is in the pair. In other words, as we mentioned before, intra-class distance information might be partially lost in FEC.
On the other side, our model performs the best compared with baseline models. It not only has the ability to synthesize Max- and Min-level samples, but also be capable of synthesizing recognizable Median-level speech samples.

\vspace{-0.3cm}

\subsection{Emotion Expressiveness}
\vspace{-0.2cm}
In this section, we conduct preference tests to evaluate whether models can express clear emotions. Since we don't consider intensity here, we only use synthesized samples with median-level intensity. 
For each utterance, we synthesize three median-level speech samples with our and two baseline models, respectively. Subjects are asked to select the one that conveys more clear emotion. If there is no detectable difference, they should choose "Same".

\begin{figure}[t]
  \centering
  \includegraphics[width=1.0\linewidth]{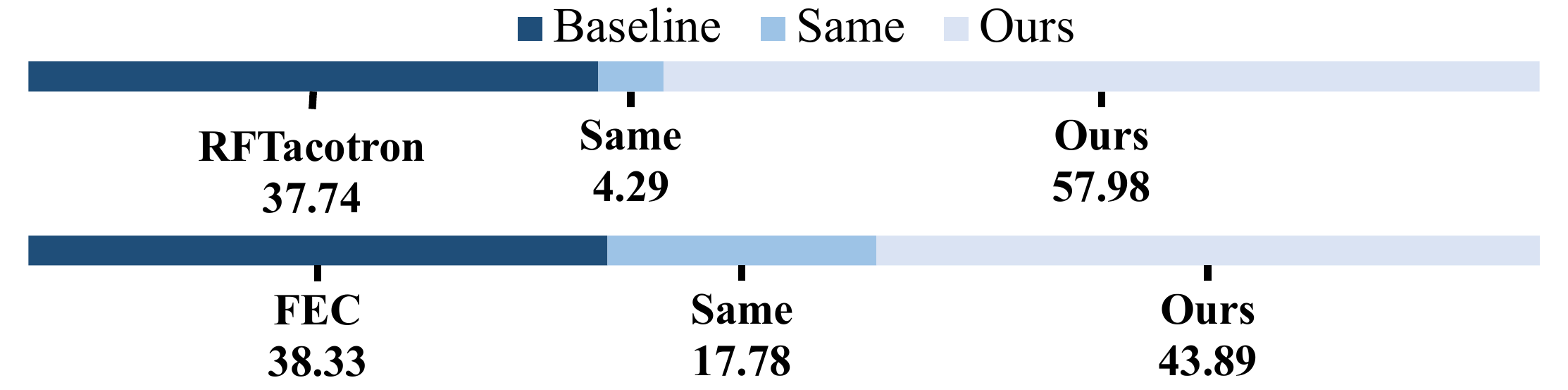}
  \caption{Preference test for emotion expressiveness.}
  \label{preference_fig}
  \vspace{-0.4cm}
\end{figure}

As we can see from the results (Fig. \ref{preference_fig}), our model significantly outperforms RFTacotron and subjects are barely confused, which suggests that our model's emotion expression is more clear than RFTacotron's.
FEC also performs well on emotion expressiveness, but our model is preferred despite the same FastSpeech2 is used for both, which implies the benefit is caused by the intensity representation of our Rank model.

\vspace{-0.3cm}
\subsection{Quality and Naturalness Evaluation}
\vspace{-0.2cm}
We further evaluate quality and naturalness of synthesized speech samples. Objective measurement Mean Cepstral Distortion (MCD) \cite{Kubichek1993MelcepstralDM}, and subjective measurement Mean Opinion Score (MOS) are conducted for this evaluation. 

Since we only focus on quality and naturalness here, and to be able to compare with ground truth speech, manual intensity labels are not used in this experiment. For FEC and our model, intensity representations are provided by their individual Rank models given ground truth speech. For RFTacotron, we use ground truth speech samples as references.

We report MCD and MOS results in Tab. \ref{mcd}. 
According to MCD results, both FEC and our model outperform RFTacotron largely, this might be because: 
1) as opposed to transferring emotion from a reference, directly assigning intensity representations is easier for the model to generate good quality speech.
2) FastSpeech2 requires less data then Tacotron2 for a high quality result.
Despite using the same FastSpeech2, the MCD of our model is slightly better than FEC. This is because our intensity representations might also bring benefits for high-quality synthesis. 
MOS scores (with 95\% confidence intervals) reveal a similar phenomenon, where FEC and our model surpass RFTacotron greatly, while our model is slightly better than FEC.

\vspace{-0.3cm}
\section{Conclusion}
\vspace{-0.2cm}
In this paper, we propose a fine-grained controllable emotional TTS, based on a novel Rank model.
The Rank model captures both inter- and intra-class distance information, and thus is able to produce meaningful intensity representations.
We conduct subjective and objective tests to evaluate our model, the experimental results show that our model surpasses two state-of-the-art baselines in intensity controllability, emotion expressiveness and naturalness.



\vfill\pagebreak

\bibliographystyle{IEEEbib}
\bibliography{myrefs}

\begin{thebibliography}{10}

\bibitem{Sotelo2017Char2WavES}
Jose M.~R. Sotelo, Soroush Mehri, Kundan Kumar, Jo{\~a}o~Felipe Santos, Kyle
  Kastner, Aaron~C. Courville, and Yoshua Bengio,
\newblock ``Char2wav: End-to-end speech synthesis,''
\newblock in {\em ICLR}, 2017.

\bibitem{Shen2018NaturalTS}
Jonathan Shen, Ruoming Pang, Ron~J. Weiss, Mike Schuster, Navdeep Jaitly,
  Zongheng Yang, Z.~Chen, Yu~Zhang, Yuxuan Wang, R.~J. Skerry-Ryan, Rif~A.
  Saurous, Yannis Agiomyrgiannakis, and Yonghui Wu,
\newblock ``Natural tts synthesis by conditioning wavenet on mel spectrogram
  predictions,''
\newblock {\em 2018 IEEE International Conference on Acoustics, Speech and
  Signal Processing (ICASSP)}, pp. 4779--4783, 2018.

\bibitem{Chung2021ReinforceAlignerRA}
Hyunseung Chung, Sang-Hoon Lee, and Seong-Whan Lee,
\newblock ``Reinforce-aligner: Reinforcement alignment search for robust
  end-to-end text-to-speech,''
\newblock in {\em Interspeech}, 2021.

\bibitem{Ren2021FastSpeech2F}
Yi~Ren, Chenxu Hu, Xu~Tan, Tao Qin, Sheng Zhao, Zhou Zhao, and Tie-Yan Liu,
\newblock ``Fastspeech 2: Fast and high-quality end-to-end text to speech,''
\newblock {\em ArXiv}, vol. abs/2006.04558, 2021.

\bibitem{Tan2021ASO}
Xu~Tan, Tao Qin, Frank~K. Soong, and Tie-Yan Liu,
\newblock ``A survey on neural speech synthesis,''
\newblock {\em ArXiv}, vol. abs/2106.15561, 2021.

\bibitem{Lee2017EmotionalEN}
Younggun Lee, Azam Rabiee, and Soo-Young Lee,
\newblock ``Emotional end-to-end neural speech synthesizer,''
\newblock {\em ArXiv}, vol. abs/1711.05447, 2017.

\bibitem{Neekhara2021ExpressiveNV}
Paarth Neekhara, Shehzeen~Samarah Hussain, Shlomo Dubnov, Farinaz Koushanfar,
  and Julian McAuley,
\newblock ``Expressive neural voice cloning,''
\newblock {\em ArXiv}, vol. abs/2102.00151, 2021.

\bibitem{Wang2018StyleTU}
Yuxuan Wang, Daisy Stanton, Yu~Zhang, R.~J. Skerry-Ryan, Eric Battenberg, Joel
  Shor, Ying Xiao, Fei Ren, Ye~Jia, and Rif~A. Saurous,
\newblock ``Style tokens: Unsupervised style modeling, control and transfer in
  end-to-end speech synthesis,''
\newblock in {\em ICML}, 2018.

\bibitem{Lee2019RobustAF}
Younggun Lee and Taesu Kim,
\newblock ``Robust and fine-grained prosody control of end-to-end speech
  synthesis,''
\newblock {\em ICASSP 2019 - 2019 IEEE International Conference on Acoustics,
  Speech and Signal Processing (ICASSP)}, pp. 5911--5915, 2019.

\bibitem{Zhu2019ControllingES}
Xiaolian Zhu, Shan Yang, Geng Yang, and Lei Xie,
\newblock ``Controlling emotion strength with relative attribute for end-to-end
  speech synthesis,''
\newblock {\em 2019 IEEE Automatic Speech Recognition and Understanding
  Workshop (ASRU)}, pp. 192--199, 2019.

\bibitem{lei2021fine}
Yi~Lei, Shan Yang, and Lei Xie,
\newblock ``Fine-grained emotion strength transfer, control and prediction for
  emotional speech synthesis,''
\newblock in {\em 2021 IEEE Spoken Language Technology Workshop (SLT)}. IEEE,
  2021, pp. 423--430.

\bibitem{lei2022msemotts}
Yi~Lei, Shan Yang, Xinsheng Wang, and Lei Xie,
\newblock ``Msemotts: Multi-scale emotion transfer, prediction, and control for
  emotional speech synthesis,''
\newblock {\em IEEE/ACM Transactions on Audio, Speech, and Language
  Processing}, vol. 30, pp. 853--864, 2022.

\bibitem{schnell2022controllability}
Bastian Schnell,
\newblock ``Controllability and interpretability in affective speech
  synthesis,''
\newblock Tech. {R}ep., EPFL, 2022.

\bibitem{Zhang2018mixupBE}
Hongyi Zhang, Moustapha Ciss{\'e}, Yann Dauphin, and David Lopez-Paz,
\newblock ``mixup: Beyond empirical risk minimization,''
\newblock {\em ArXiv}, vol. abs/1710.09412, 2018.

\bibitem{wang2022zero}
Shijun Wang and Damian Borth,
\newblock ``Zero-shot voice conversion via self-supervised prosody
  representation learning,''
\newblock in {\em 2022 International Joint Conference on Neural Networks
  (IJCNN)}. IEEE, 2022, pp. 01--08.

\bibitem{McAuliffe2017MontrealFA}
Michael McAuliffe, Michaela Socolof, Sarah Mihuc, Michael Wagner, and Morgan
  Sonderegger,
\newblock ``Montreal forced aligner: Trainable text-speech alignment using
  kaldi,''
\newblock in {\em INTERSPEECH}, 2017.

\bibitem{Adigwe2018TheEV}
Adaeze Adigwe, No{\'e} Tits, Kevin~El Haddad, Sarah Ostadabbas, and Thierry
  Dutoit,
\newblock ``The emotional voices database: Towards controlling the emotion
  dimension in voice generation systems,''
\newblock {\em ArXiv}, vol. abs/1806.09514, 2018.

\bibitem{Yamamoto2020ParallelWA}
Ryuichi Yamamoto, Eunwoo Song, and Jae-Min Kim,
\newblock ``Parallel wavegan: A fast waveform generation model based on
  generative adversarial networks with multi-resolution spectrogram,''
\newblock {\em ICASSP 2020 - 2020 IEEE International Conference on Acoustics,
  Speech and Signal Processing (ICASSP)}, pp. 6199--6203, 2020.

\bibitem{Sivaprasad2021EmotionalPC}
Sarath Sivaprasad, Saiteja Kosgi, and Vineet Gandhi,
\newblock ``Emotional prosody control for speech generation,''
\newblock in {\em Interspeech}, 2021.

\bibitem{Kubichek1993MelcepstralDM}
Robert~F. Kubichek,
\newblock ``Mel-cepstral distance measure for objective speech quality
  assessment,''
\newblock {\em Proceedings of IEEE Pacific Rim Conference on Communications
  Computers and Signal Processing}, vol. 1, pp. 125--128 vol.1, 1993.

\end{thebibliography}

\end{document}